\documentclass{iopart}
\usepackage{amssymb}
\usepackage{graphicx}
\usepackage[usenames,dvipsnames]{xcolor}
\usepackage{bm}

\newcommand{\eqref}[1]{(\ref{#1})}

\usepackage{hyperref}
\hypersetup{%
  pdftitle={Lattice scars: Surviving in an open discrete billiard}
  anchorcolor=blue,
  colorlinks=true,
  citecolor=blue,
  urlcolor=blue,
  linkcolor=blue}

\begin{document}

\title{
Lattice scars: Surviving in an open discrete billiard}

\author{%
 V\'ictor Fern\'andez-Hurtado$^1$,
 Jordi Mur-Petit$^2$,
 Juan Jos\'e Garc\'ia-Ripoll$^1$
 and
 Rafael A. Molina$^2$}
\address{$^1$ Instituto de F\'isica Fundamental, IFF-CSIC, Serrano 113b, Madrid 28006, Spain}
\address{$^2$ Instituto de Estructura de la Materia, IEM-CSIC, Serrano 123, Madrid 28006, Spain}
\ead{jordi.mur@csic.es}


\begin{abstract}
We study 
quantum systems on a discrete bounded lattice (lattice billiards).
The statistical properties of their spectra show universal features related to the regular or chaotic character of 
their classical continuum counterparts.
However, the decay dynamics of the open systems appear very different from the continuum case, their properties being dominated by the states in the band center.
We identify a class of states (``lattice scars'') that survive for infinite times in dissipative systems and that are degenerate at the center of the band.
We provide analytical arguments for their existence in any bipartite lattice, and give a formula to determine their number.
These states should be relevant to quantum transport in discrete systems, and we discuss how to observe them using photonic waveguides, cold atoms in optical lattices, and quantum circuits.
\end{abstract}




\submitto{\NJP}

\maketitle


\section{Introduction}

Understanding and controlling quantum transport is essential for many different quantum technologies. By now, it has become clear than different quantum systems with very different sizes and time scales follow the same
guiding principles as far as transport properties go \cite{Datta_Book,Nazarov_Book}.
One of these guiding principles is that the statistical properties of
quantum transport and quantum decay are chiefly determined by the
chaotic or regular properties of the classical analog.
One of the most remarkable achievements in classical mechanics in the last century has been the establishment that the time evolution of certain dynamical systems is {\em chaotic}, i.e., it features an extreme sensitivity to initial conditions, usually portrayed by their Lyapunov exponents, which are a measure of an exponential divergence of trajectories in phase space.
Even though the concept of trajectory no longer holds in quantum physics, {\em
  quantum chaos}, the quantum-mechanical study of classically chaotic systems, has also flourished~\cite{stockmann1999book}.
Results of quantum chaos have been particularly remarkable in the study of billiards: domains wherein a particle moves ballistically except for elastic collisions with the boundary.
One of the most surprising results in this field was the discovery by Heller~\cite{heller1984} that the probability amplitude of certain wavefunctions---called ``scarred wavefunctions'' or, simply, ``scars''---in a chaotic two-dimensional billiard is not uniform but concentrates along the trajectory of classical periodic orbits. 
This effect due to wave interference has now been observed in a number of systems, from microwaves in cavities~\cite{sridar1991,stein1992}, to electrons in quantum dots~\cite{marcus1992,akis1997}, to optical fibers~\cite{michel2012}. 
The relevance on quantum transport of quantum-chaotic effects in general, and scarred states in particular, is widely supported by theory and experimental evidence~\cite{beenakker1997rmp,alhassid2000}.

The chaotic or regular properties of the dynamics in the closed system have important consequences when the system is opened. In particular, the decay properties
of the particles inside a leaking billiard depend strongly on the system being regular or chaotic, on the presence of marginally
stable periodic orbits (\textit{bouncing balls})~\cite{bauer1990}, and on Sieber-Richter ``paired'' trajectories~\cite{richter2002,waltner2008}. 
With the development of atomic cooling and trapping techniques, beautiful experiments could be performed exploring different issues of quantum chaos~\cite{raizen2011}.
The group of Nir Davidson confined rubidium atoms to a billiard realized by rapidly scanning a blue-detuned laser beam following the shape of the desired domain~\cite{milner2001,friedman2001}. Opening a hole in the billiard, the number of atoms trapped as a function of time followed an exponential decay for chaotic domains, and a power-law decay for domains supporting stable trajectories~\cite{milner2001}. They also showed the controlled appearance of islands of stability when the walls of chaotic billiards are softened~\cite{kaplan2001} in agreement with theoretical arguments~\cite{alt1996}.
However, no scarred states were observed.

\begin{figure}[t]
  \centering
  \includegraphics[width=\linewidth]{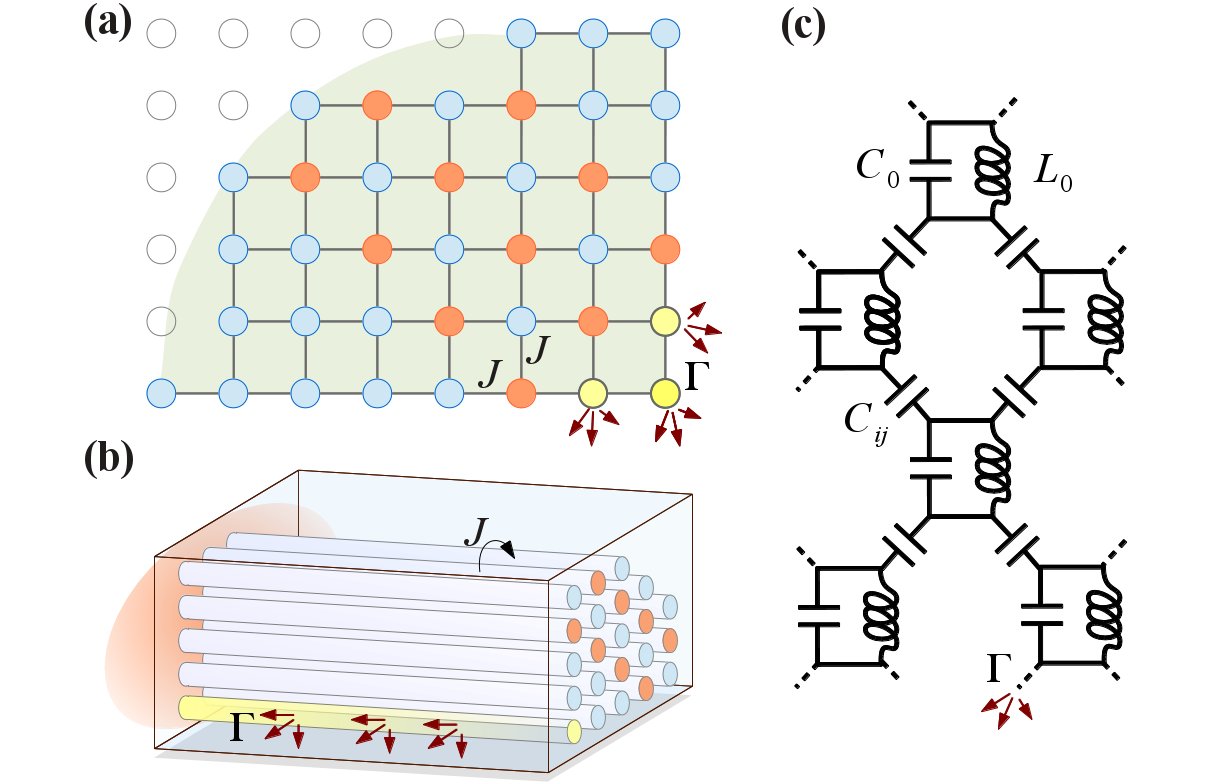}
  \caption{
     From a continuous region (a stadium, a rectangle, etc.), we obtain a {\em discrete billiard} by selecting only the sites (circles) that are inside it (shaded region).
     Particles are allowed to hop between sites with probability $J$, and there is a sink of particles (decay rate $\Gamma$) at a corner of the lattice.
     This model can be implemented using optical lattices (a), coupled photonic waveguides (b) or coupled superconducting microwave resonators (c).
     In the first case, the sink can be implemented with a focused, resonant laser. For coupled waveguides, it is a guide with losses, while for (c) the loss comes from a resistor or a semi-infinite transmission line coupled to a few resonators.
   }
  \label{fig:sketch}
\end{figure}

In this work, we study regular and chaotic billiards where the particle motion is restricted to a square lattice of discrete points. This model is adequate to describe several systems, including ultra-cold atoms trapped in optical lattices~\cite{bloch2012nphysrev,lewenstein2012book}, cf.~Fig.~\ref{fig:sketch}a, as well as the propagation of light along photonic waveguides~\cite{politi2008,obrien2009}, Fig.~\ref{fig:sketch}b.
We consider billiards that in a continuum description feature regular and chaotic properties. By studying the statistics of their energy levels, we show that these behaviors are also present for the discrete case. Furthermore, we study the quantum dynamics in dissipative billiards with a leak localized on the border, and show that the population in {\em both} kinds of systems follows a similar trend: an initial exponential decay, followed by a power-law decay, until on occasions a final non-zero population is trapped in the system. We explain this unexpected behavior by the appearance of ``lattice scars'': scarred wavefunctions supported on the lattice structure and whose energy is at the band center, $E=0$.
Our numerical findings are supported with analytical arguments, which lay the necessary conditions for the appearance of these states, thus pointing a route for controlling the dissipation in finite lattice systems.
Finally, we discuss the observability of this effect in several different atomic, photonic, and solid-state setups.

\section{Energy statistics}
We start by computing the eigenvalues $E_n$ and eigenfunctions $\psi_n$ of a lattice Hamiltonian 
\begin{equation}
  H = -\sum_{\langle l,m\rangle} J_{lm}c^{\dagger}_l c_m
  \label{eq:ham}
\end{equation}
where $J_{lm}~(l,m=1,\ldots,N)$ is the hopping amplitude from site $m$ to site $l$, $c_m~(c^{\dagger}_m)$ destroys (creates) a particle at site $m$, and the sum runs over all pairs of nearest neighbors of the $N$-points lattice. The topology of the billiard is hence encoded in the hopping amplitudes or, equivalently, on the set of neighbors of a given site.

We calculate the eigenvalues by exact diagonalization. The Hamiltonian presents chiral symmetry meaning that the Schr\"odinger equation can be written as 
\begin{eqnarray}\label{chiralEq}
H\Psi=
 \left( \begin{array}{cc} 0 & C \\ C^T & 0 \end{array} \right)
 \left( \begin{array}{c} \Psi_A \\ \Psi_B \end{array} \right)
= E \left( \begin{array}{c} \Psi_A \\ \Psi_B \end{array} \right),
\end{eqnarray}
with $A$ and $B$ representing the two sublattices in which the square lattice can be divided. Sites in the $A$ sublattice only connect with sites in the $B$ sublattice and vice versa. This bipartite property of the square lattice translates into a symmetry of the eigenenergies around the band center $E=0$.

Following the standard procedure and taking into account the symmetries in the spectrum, we unfold the set of eigenenergies into
$s_n=(E_{n+1}-E_n)/\langle E_{n+1}-E_n \rangle$, where the brackets $\langle
\cdot \rangle$ denote a local average. We have used different unfolding procedures and checked that the spacing distribution, $P(s)$, obtained is
the same, including a local unfolding with different energy windows~\cite{haake}, as well as using a smooth functional form that takes into account the logarithmic divergence of the density of states at the band center.
The normalized level spacing distribution, $P(s)$, for a continuum regular billiard follows a Poisson distribution, $P_{\mathrm{P}}(s) = \exp(-s)$~\cite{berry1977}, while for chaotic billiards it follows the Wigner surmise,
\begin{equation}
  P_{\mathrm{W}}(s) = \frac{\pi}{2}se^{-\pi s^2/4} \:,
  \label{eq:wigner}
\end{equation}
from Random Matrix Theory (RMT)~\cite{bohigas1984,stockmann1990}. 
In the case of the lattice billiards with a square lattice that we are
  studying the proper Random Matrix Ensemble is the chiral Gaussian Orthogonal
  Ensemble (ch-GOE, or BD I in the Cartan classification of symmetric spaces) for systems with time reversal and chiral symmetries~\cite{altland97}. However, besides the symmetry of the spectrum around
the band center, the statistical spectral properties, including the $P(s)$,
are the same as for the usual GOE.

\begin{figure}[t]
  \centering
  \includegraphics[width=0.8\linewidth]{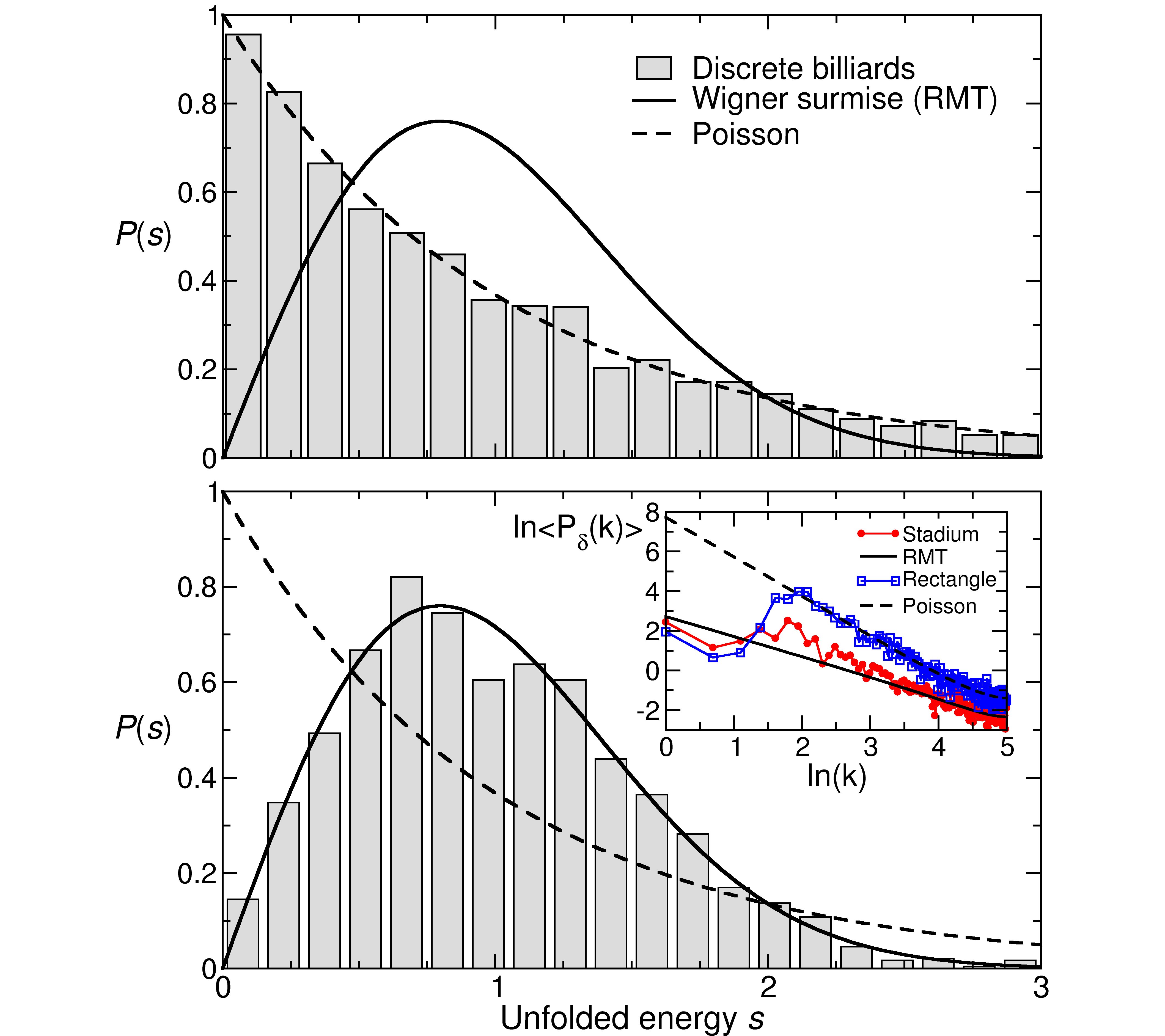}
  \caption{
  Level spacing statistics for (top) rectangle and (bottom) stadium billiards. Numerical data are plotted with bars while the lines are the theoretical predictions of the Poisson distribution (dashed) and RMT (solid). 
  Inset: long-range energy statistics averaged over 300 states close to the band center of 10 rectangular (squares) and 10 stadium (dots) billiards with similar total number of sites, and the theoretical predictions with the same line coding.}
  \label{fig:PdeS}
\end{figure}

We show in Fig.~\ref{fig:PdeS} our numerical
results for a rectangular billiard of $50\times 37$ sites [Fig~\ref{fig:PdeS}(a)] and for a desymmetrized Bunimovich stadium with a total of 5238 lattice sites [Fig.~\ref{fig:PdeS}(b)]. Similarly to the continuum case, we see that 
the former agrees well with a Poisson distribution (\textit{dashed line}), while the stadium presents a distribution in agreement with RMT (\textit{solid}). It is worth mentioning that we do not find any indication of the semi-Poisson behavior that was found for the very similar spin stadium billiard in Ref.~\cite{montangero2009}.
We have also performed a more stringent test, based on an analysis of the long-range correlations, calculated through the Power Spectrum, $P_{\delta}(k)$, of the $\delta_n$ statistics (as defined in Refs.~\cite{relano2002,faleiro2004}).
Our numerical results
are presented in the inset of Fig.~\ref{fig:PdeS}. The comparison with the theoretical expectations is very good, including the decrease in the average value of the Power Spectrum for small values of the frequency $k$, which can be understood as the effect of bouncing ball orbits~\cite{faleiro2006}.

\section{Dynamics in open systems}
Having established the static properties of the discrete rectangular and stadium billiards, we proceed now to analyze their dynamics in the presence of dissipation, which we include in the form of a leaking hole on the border of the system. We have studied the evolution of a localized wavepacket with initial momentum $\bm{p}_0$ and width $w$, described by a pure state $\psi_i(t=0) \propto \exp[-(\bm{x}_i -\bm{x}_0)^2/2w^2 - i \bm{p}_0 \cdot \bm{x}_i]$.
For a weak dissipation, the dynamics of the resulting mixed state is given by the master equation,
\begin{equation}
  \frac{\partial\rho}{\partial t}
  =
  -\frac{i}{\hbar}[H,\rho]
  +
  \sum_k \frac{\gamma_k}{2\hbar} \left(
  	2 c^{\dagger}_k	\rho c_k - c_k c^{\dagger}_k \rho - \rho c_k c^{\dagger}_k
  	\right) \:.
 \label{eq:master}
\end{equation}
Here, $H$ is given by Eq.~\eqref{eq:ham} while $\gamma_k$ describes the loss
rate: $\gamma_k=\Gamma$ within the leak located on the billiard boundary, and
$\gamma_k=0$ otherwise. This is equivalent to the evolution under an effective non-Hermitian Hamiltonian with imaginary on-site energies $\gamma_k$.
We used a value $\Gamma=2$ (in units of nearest-neighbor hopping $J$) for the decay rate, and a leak radius $\sigma=2$ (in units of the lattice constant).
We have verified that using a square-well or Gaussian profile for the leak does not substantially modify our findings.

\begin{figure}[t]
  \centering
  \includegraphics[width=\linewidth]{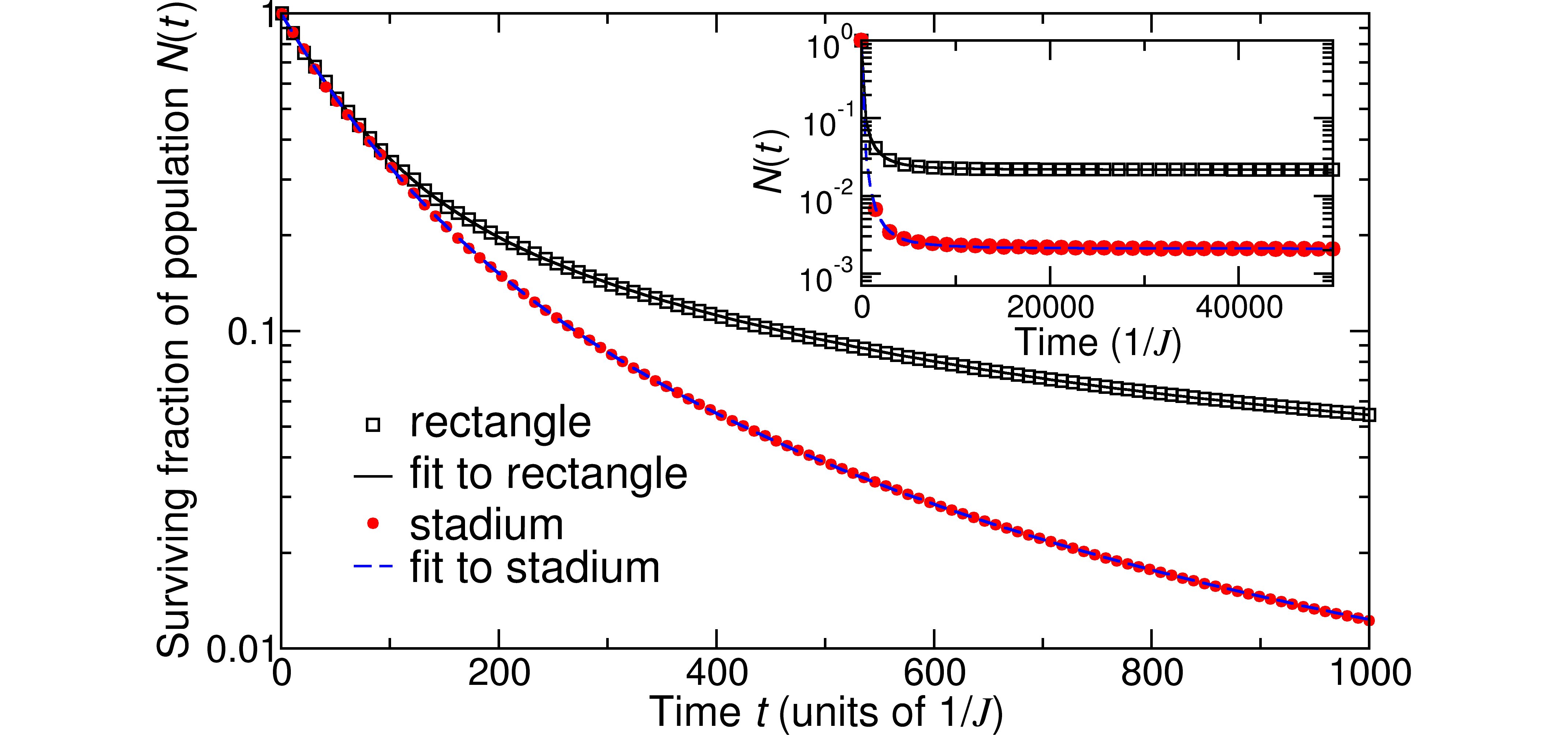}
  \caption{
  Fraction of population trapped in the lattice for the rectangle (top curve) and stadium (bottom curve), averaged over the position of the leak around the billiard. Symbols are simulation data while lines stand for least-squares fits to Eq.~\eqref{eq:NdeT}. Inset: behavior for long times.}
  \label{fig:NdeT}
\end{figure}

The number of particles remaining in the system after a time $t$ is 
$N(t) = \sum_k \mathrm{Tr}(\rho(t) c^{\dagger}_k c_k)$.
The average of $N(t)$ over all possible positions of the hole, and over a range of initial momenta $\bm{p}_0$ is shown in Fig.~\ref{fig:NdeT}.
For classical systems, one expects very different population dynamics for the two billiards~\cite{alt1996,bauer1990,dettmann2009}: a rapid exponential decay for the chaotic one, and a power-law decay for the regular one. For quantum systems, unless there is a large number of decay channels or holes, a purely algebraic decay is expected~\cite{alt1995,alt1996}. These predictions have been confirmed in previous experiments in a large variety of continuous systems, from microwaves billiards~\cite{alt1995} to cold atoms in optics billiards~\cite{friedman2001}.  Here, we observe two features that strikingly contradict these expectations: (i) the population dynamics is similar for {\em both} discrete billiards, and (ii) there is a fraction of population that remains trapped for arbitrarily long times. Indeed, $N(t)$ decays rapidly at short times $tJ\lesssim 1000$, then it levels off, and finally saturates for $tJ\gtrsim 10^4$ [cf.~Fig.~\ref{fig:NdeT}, inset]. The numerical data is accurately fitted by the formula (
compare Eq.~(1) in Ref.~\cite{alt1996})
\begin{equation}
  N(t) =
  {\cal E} \exp(-\lambda t)
  + {\cal A} (1+\alpha t)^{-\beta}
  + {\cal S} \:.
  \label{eq:NdeT}
\end{equation}
For a given initial wavepacket and position of the hole, this can be rationalized in terms of the decomposition of $\psi_i(0)$ over the eigenstates of the closed billiard with rapid (exponential) decay for short times, ${\cal E}$, those with algebraic decay, ${\cal A}$, and those that survive the presence of the leak for $t\gtrsim 10^4$, ${\cal S}=N(t=0)-{\cal E - A}$.
The rapidly decaying eigenstates correspond to those that overlap the site where the leak opens, or to trajectories of the wavepacket that reach the hole after only a few bounces off the walls;
classically, this can be expected to be most relevant for chaotic systems, where (almost) any initial trajectory will quickly approach the leak.
Algebraic decay is associated with orbits that go through many bounces before leaking out~\cite{dettmann2009}.
We do not expect Sieber-Richter paired trajectories~\cite{richter2002,waltner2008} to be relevant here, as discretized systems do not support exponentially close pairs of trajectories.

Following this idea, we have performed a quantitative analysis of the eigenenergies, $E_n^{\mathrm{open}} = \varepsilon_n + i\Gamma_n$, of the non-Hermitian Hamiltonian with imaginary on-site energies, for rectangular and stadium billiards. In both cases, we find that the widths $\Gamma_n$ can be divided into three sets:
(i) a very small number of states (2$-$8) with large imaginary parts, $\Gamma_n \geq 10^{-1}$, which we expect to decay for times $\leq 10^1$. (ii) A large fraction ($\gtrsim95\%$) of states with $\Gamma_n \approx 10^{-4}-10^{-1}$ which decay slowly and whose widths follow, in the chaotic billiard, a Porter-Thomas distribution as RMT predicts~\cite{stockmann1999book} (see~\ref{app:porter}).
Finally, (iii) a small number of states with very small decay rates; among these, a few $\Gamma_n$ are numerically equivalent to zero.


%
\begin{figure}[t]
  \centering
  \includegraphics[width=\linewidth]{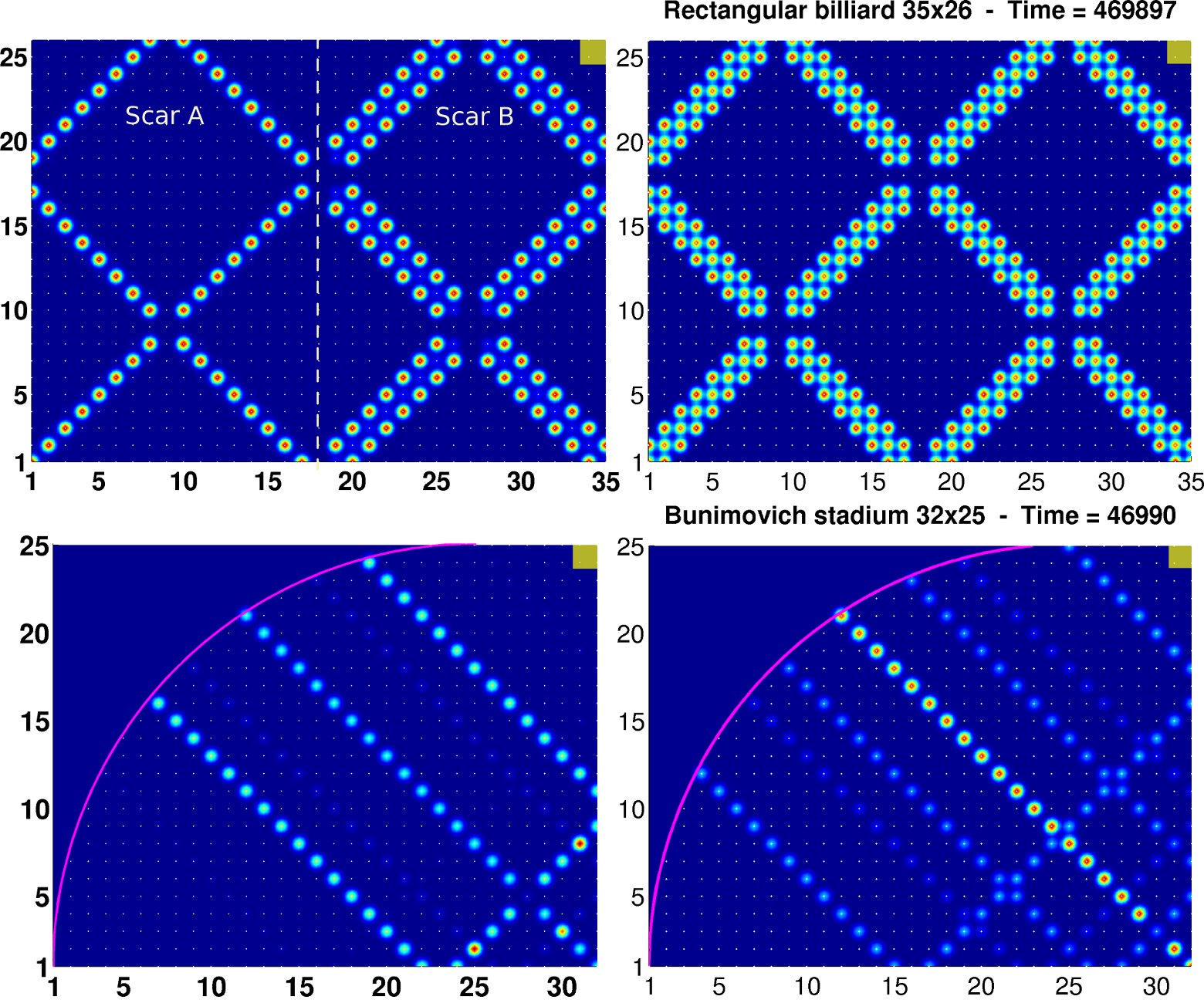}
  \caption{
     Probability density of some states in lattices with a leak at the top-right corner (yellow square) of a rectangular billiard with $M=35,N=26$ (top row) and a stadium with $M=32,N=25$ (bottom).
     Left column: lattice scars, i.e., $E_n^{\mathrm{open}}=0$ eigenstates of the non-Hermitian Hamiltonian: there are two for the rectangle (each localized on a different sublattice, shown on the same figure as they are symmetric upon reflection on the central (dashed) line), and four for the stadium (of which we show one).
     Right column: snapshots of the dynamical evolution with Eq.~\eqref{eq:master} at the indicated times.
     Lower density is indicated by blue (dark grey) and higher by lighter colors; maximum density is at red spots.
     Small white dots point the lattice sites, and the purple line is the circular edge of the stadium. See~\cite{epaps}.}
  \label{fig:scars}
\end{figure}

\section{Lattice scars}
The probability density of one of these non-decaying eigenstates, $|\psi_n^{\mathrm{open}}|^2$, in the rectangle (stadium) is shown on the left panel of Fig.~\ref{fig:scars} (top, resp.\ bottom) when the leak is at the top-right corner of the billiard. The long lifetime of these states is quickly understood by noting their vanishing densities at the hole position. Their spatial distribution on the billiard, however, is far from a ``bouncing ball'' orbit~\cite{alt1995,alt1996,dettmann2009,loeck2012}. This is due to the lattice constraint in~$J_{lm}$ that only allows to hop from one site to its nearest neighbor. This, together with the geometry of the square lattice, amounts to the system being bipartite on two disjoint sublattices, $A,B$, as mentioned in the discussion about spectral statistics. Application of a theorem by Inui {\em et al.}~\cite{inui1994} implies then that the closed system has $n$ solutions to the Schr\"odinger equation at the band center, $E=0$, which vanish on one of the sublattices, say 
$\psi_{k \in B}=0$. Here, $n$ is the number of sites on the occupied lattice (here, $N_A$) minus the rank, $r$, of the matrix $J_{lm}$~\cite{inui1994}, i.e., $r+n=N_A$.
Once we open the system, some of these degenerate states in the middle of the band stay at $E=0$, while the others acquire a purely imaginary eigenvalue with a large width. The number of the latter is given directly by the number of $A$ sites overlapping the leak. These large-width states dominate the decay at short times.
States remaining at $(\varepsilon_n=0,\Gamma_n=0)$, on the other hand, will dwell in the billiard for very long times. We refer to them as {\em lattice scars}.
%

\begin{figure}[t]
  \centering
  \includegraphics[width=\linewidth]{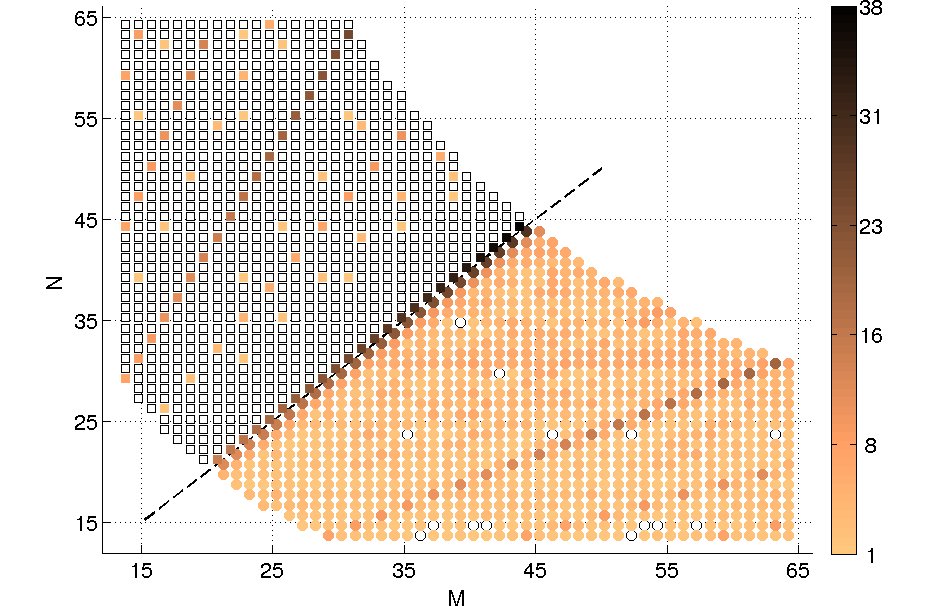}
  \caption{
   Number of lattice scars for a given billiard.
   Squares 
   represent data for $M\times N$ rectangular billiards; 
   circles show data for stadia with circular ends of radius $N$ and straight edges of length $M-N$.
   The dashed line separates the data sets, which have been slightly shifted to ease visualization of the points along $M=N$.
   The number of lattice scars is indicated by color darkness, with black the maximum; empty symbols correspond to billiards with no lattice scars.}
  \label{fig:scar_geometry}
\end{figure}

We have calculated the number of lattice scars for a wide range of system sizes, as shown in Fig.~\ref{fig:scar_geometry}.
We see that most rectangles feature no lattice scars: they appear only when $M/N\approx q,~q\in\mathbb{Z}$. In contrast, almost all stadia have at least one lattice scar, with a larger number when $M\approx2N$, a trace itself of the embedded rectangle~\footnote{Note that for $M=N$ a Bunimovich stadium actually corresponds to a circle of radius $M$, which is classically a regular billiard.}.

The spatial distribution of the bound $E^{\mathrm{open}}=0$ states will hence reside in sites $k \in A$, which on a square lattice are linked by $45^{\circ}$ lines. This requirement, besides the boundary conditions appropriate to each billiard, which restrict the allowed ``bounces" off the walls, results in trapped eigenfunctions such as those in Fig.~\ref{fig:scars}. Indeed, we have independently checked that {\em all} states with $\Gamma_n=0$ do have $\varepsilon_n = 0$. Moreover, we have also verified the prediction derived from the theorem in~\cite{inui1994} that the number of states with
$\varepsilon_n=\Gamma_n=0$
equals $n$ as defined above.

%

From the preceding arguments, we conclude that an initial state on a dissipative discrete billiard will evolve until at times $t\gg 1/J$, all probability amplitude is concentrated on a (superposition of) lattice scarred state(s), i.e., \textit{dissipation selects this class of eigenstates}, removing all the other components of the initial wavepacket. This prediction is confirmed by looking at the probability density of an initial state after numerically evolving it according to Eq.~\eqref{eq:master}, see the right panels in Fig.~\ref{fig:scars}: the resemblance of these with the eigenstate probability densities on the left panels is evident~\cite{epaps} and we verified that these stationary wavefunctions are pure superpositions of states with $\varepsilon_n=\Gamma_n=0$.


\section{Physical implementations}
\subsection{Photonic waveguides}
Paired trajectories in the sense of~\cite{richter2002} strongly affect the conductance through quantum dots~\cite{richter2002,alhassid2000}. Analogously, we expect lattice scars to influence transport through discrete systems.
As a first example, this effect can be studied with infrared or visible light in photonic lattices, which are optical circuits with tens of waveguides imprinted on a substrate using a laser~\cite{fleischer2003,obrien2009,krimer2011}. These waveguides can be arranged on parallel rows forming a square lattice which are evanescently coupled, i.e., light may tunnel between neighboring guides, cf.\ Fig.~\ref{fig:sketch}b. Doping one or more of the waveguides, or coupling them to outgoing lines, would result in a sink realizing the desired amount of dissipation. The paraxial propagation of light in this setup is described by Eq.~\eqref{eq:master}, the final population distribution corresponding to the distribution of light at the substrate's end.

Typical numbers for such a system are a guide length $L=1-10$~cm, width $W \approx 200~\mu$m, and inter-guide distance of $d \approx 20~\mu$m, resulting in a coupling
$J \approx 1-10~\mathrm{cm}^{-1}$~\cite{szameit2010jpb,garanovich2012,corrielli2013}.
An initial wavepacket corresponds in this system to a spatial intensity distribution over the waveguides on the $z=0$ plane, $|\psi(x,y;z=0)|^2$, which then propagates along the guides' length, $z$, that amounts to the time variable in our simulations.
Then, the distance, $z_{\mathrm{scar}}$, after which the intensity profile has taken the shape of the lattice scar equals the characteristic time, $t_{\mathrm{scar}}$, required for scar appearance as seen in the simulations.
For systems with $\sim 10\times10$ guides (as afforded by $W$), we find $z_{\mathrm{scar}} \equiv t_{\mathrm{scar}}\approx10^{3} J^{-1}$. Hence, to get $z_{\mathrm{scar}} < L$ one requires $J \gtrsim 10^3 L^{-1} \gtrsim 10^2~\mathrm{cm}^{-1}$.
For waveguides written on fused silica using fs laser pulses~\cite{szameit2010jpb}, this amounts to creating a refractive index variation
$\Delta n = J \lambda/2\pi \approx 1\times 10^{-3}$ for visible light, which is well within present capabilities~\cite{szameit2010jpb}. 
Under these conditions, for $z_{\mathrm{scar}} \lesssim z \leq L$ the light beam propagates with a constant intensity profile, similarly as in lattice solitons~\cite{christo2003} but without nonlinear effects.

\subsection{Cold atoms in optical lattices}
Our predictions can also be investigated using cold atoms trapped in optical lattices~\cite{ng2009} (Fig.~\ref{fig:sketch}a), where single-site resolution for preparation and measurement has already been demonstrated in several labs~\cite{bakr2009,wurtz2009,sherson2010}.
Dissipation here would be realized via a focused blue-detuned laser beam, which can be pointed either on the system boundary or even inside the billiard.
A major challenge in these systems is to produce a lattice with a customized boundary, a task that can be achieved thanks to the improved optics in recent experiments, which allows projecting arbitrary optical potentials onto the trapping plane~\cite{bakr2009}.
Detection of quantum transport modifications due to the lattice topology in these atomic systems would contrast the observations in graphene, where weak localization is strongly suppressed~\cite{morozov2006,dassarma2011rmp}.

Approximate time and length scales in these setups are an inter-site separation $d\approx\lambda/2 \approx 500$~nm ($\lambda$ being the optical wavelength) and hopping energy $J \approx \hbar \times (1-100)$~kHz~\cite{bloch2012nphysrev,lewenstein2012book}, leading to $t_{\mathrm{scar}} \approx 10^3 \hbar/J \approx 10\,\mathrm{ms}-1$~s, which lies within the typical lifetime of these systems.

\subsection{Superconducting microwave circuits}
Finally, the same ideas can be studied using microwave quantum optics. Inspired by recent designs of coupled harmonic oscillators~\cite{houck2012}, we suggest creating a lattice of capacitively coupled microwave LC resonators, Fig.~\ref{fig:sketch}c.
When the capacitive coupling is weaker than the on-site energy, the rotating-wave approximation applies~\cite{Peropadre2013} and the hopping of microwave photons in the array is described once more by Eq~\eqref{eq:master}.
The leak can be introduced using either resistive elements or outgoing wires that extract energy from a few sites. The distribution of energy can be measured using a probe antenna that is moved over the circuit to scan the electromagnetic field. The timescale of such experiments is much faster than in the atomic case. Both the oscillator energy and the coupling between oscillators can be within the range of GHz to tens of MHz, to ensure that we are far above the typical decoherence times of the cavities. Assuming that the dissipation has the same rate as in the other setups, once more the observation timescale of the decay $t_{\mathrm{scar}}\sim 1\,{\mathrm{ns}}-0.1\,\mu\mathrm{s}$, which would allow for a fast preparation of the state while being able to monitor the decay of the electromagnetic field.


\section{Conclusions}
In summary, we studied two quantum billiards on a bipartite lattice: a rectangle and the Bunimovich stadium. We have shown that their level statistics agree with those of a regular and a chaotic billiard, respectively. However, the dynamics of a wavepacket on the open billiards turns out to be rather similar for both cases, and presents a number of, to the best of our knowledge, so-far unnoticed features. The most remarkable is the appearance of lattice scars: states at the band center whose probability density collapses around spatially-concentrated orbits that live on only one of the two sublattices. This allows them to survive the presence of localized decay channels for very long times. We determined analytically the surviving population in terms of the lattice geometry and hoppings. Finally, we  discussed three experimental setups, within current capabilities, to test our predictions.

Propagation through periodic lattices is a subject of interest in fields as diverse as  biological molecules~\cite{plenio2008,seeman2005}, optical waveguides~\cite{fleischer2003,obrien2009}, nanophysics~\cite{mello2004,datta2005} and cold-atom systems~\cite{bloch2012nphysrev,lewenstein2012book},
and we expect that this work will enable new perspectives in the study and control of quantum dynamics in classically-chaotic regimes.
These results should also be relevant in
quantum simulations of interacting systems~\cite{bloch2012nphysrev}, quantum walks~\cite{perets2008,karski2009}, and quantum-enhanced computational techniques such as boson sampling~\cite{broome2013,spring2013,tillmann2013,crespi2013}.

\ack
We thank V.\ A.\ Gopar for bringing Ref.~\cite{inui1994} to our attention.
This work has been funded by Spanish Government projects FIS2012-33022, FIS2009-07277, FIS2012-34479, CAM research consortium QUITEMAD (S2009-ESP-1594),
COST Action IOTA (MP1001), 
EU Programme PROMISCE, ESF Programme POLATOM, and the JAE-Doc and JAE-Intro programs (CSIC and European Social Fund).

\appendix


\section{Statistics of eigenstate widths: Porter-Thomas distribution}
\label{app:porter}
\setcounter{section}{1} 

\begin{figure}[bt]
  \centering
  \includegraphics[width=0.8\linewidth]{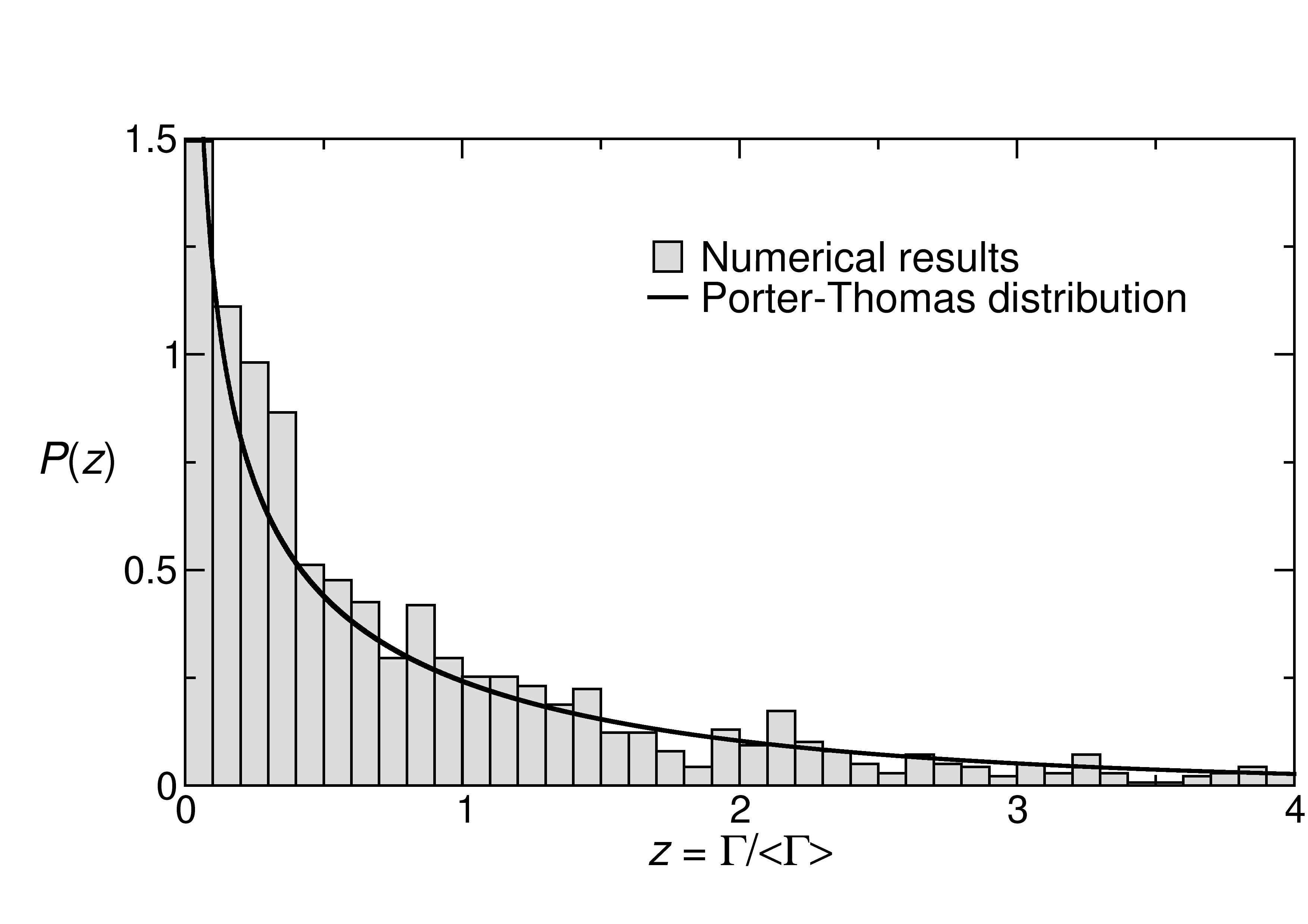}
  \caption{
  Comparison of the distribution of widths for the stadium billiard (bars) and the Porter-Thomas distribution, Eq.~\eqref{eq:PT} (solid line).}
  \label{fig:PT}
\end{figure}

Porter and Thomas derived the distribution of partial widths of the resonances
for an open quantum system assuming a Gaussian distribution of the eigenstates
amplitudes. The Porter-Thomas distribution can also be derived from Random
Matrix Theory~\cite{Porter56,Brody81}.
\begin{equation}
 P_{\mathrm{PT}}(z)=\frac{1}{\sqrt{2 \pi z}} \exp{(-z/2)},
 \label{eq:PT}
\end{equation}
where $z=\Gamma/\left<\Gamma\right>$. Eigenstates following the random wave model, conjectured to be valid for the statistical properties of chaotic quantum system by Berry~\cite{Berry77}, fulfill this property. We have calculated the distribution of the widths (imaginary parts of the energy) of the eigenstates corresponding to the effective Hamiltonian of an opened stadium lattice billiard. In order to make a fair comparison, the average width value $\left<\Gamma\right>$ was calculated averaging in a window of $12$ neighboring resonances in energy~\cite{Meredith93} and over 10 different positions of the opening for a stadium billiard of 2784 sites without taking into account the states in the band center. The results are shown in Fig.~\ref{fig:PT}.

The agreement is quite good taking into account that small divergences are expected due to a breakdown of the random wave model due to the walls and/or
discreteness of the lattice. From the results shown we can conclude that the random wave model is a good approximation from the behavior of the
wave functions outside the band center in chaotic lattice billiards. A more careful analysis of the differences between the calculated width distribution and the Porter-Thomas distribution is beyond the scope of this work.  

%
\section*{References}
\bibliographystyle{unsrt}
\bibliography{scars}

\begin{thebibliography}{10}

\bibitem{Datta_Book}
S.~Datta.
\newblock {\em Quantum Transport: {A}tom to {T}ransistor}.
\newblock Cambridge University Press, Cambridge, 2005.

\bibitem{Nazarov_Book}
Y.V. Nazarov and Y.M. Blanter.
\newblock {\em Quantum Transport: {I}ntroduction to {N}anoscience}.
\newblock Cambridge University Press, Cambridge, 2013.

\bibitem{stockmann1999book}
H.-J. St{\"o}ckmann.
\newblock {\em Quantum {C}haos: {A}n {I}ntroduction}.
\newblock Cambridge University Press, Cambridge, UK, 1999.

\bibitem{heller1984}
Eric~J. Heller.
\newblock Bound-state eigenfunctions of classically chaotic {H}amiltonian
  systems: Scars of periodic orbits.
\newblock {\em Phys. Rev. Lett.}, 53(16):1515--1518, October 1984.

\bibitem{sridar1991}
S.~Sridhar.
\newblock Experimental observation of scarred eigenfunctions of chaotic
  microwave cavities.
\newblock {\em Phys. Rev. Lett.}, 67:785--788, Aug 1991.

\bibitem{stein1992}
J.~Stein and H.-J. St\"ockmann.
\newblock Experimental determination of billiard wave functions.
\newblock {\em Phys. Rev. Lett.}, 68:2867--2870, May 1992.

\bibitem{marcus1992}
C.~M. Marcus, A.~J. Rimberg, R.~M. Westervelt, P.~F. Hopkins, and A.~C.
  Gossard.
\newblock Conductance fluctuations and chaotic scattering in ballistic
  microstructures.
\newblock {\em Phys. Rev. Lett.}, 69:506--509, Jul 1992.

\bibitem{akis1997}
R.~Akis, D.~K. Ferry, and J.~P. Bird.
\newblock Wave function scarring effects in open stadium shaped quantum dots.
\newblock {\em Phys. Rev. Lett.}, 79:123--126, Jul 1997.

\bibitem{michel2012}
C.~Michel, S.~Tascu, V.~Doya, P.~Aschi\'eri, W.~Blanc, O.~Legrand, and
  F.~Mortessagne.
\newblock Experimental phase-space-based optical amplification of scar modes.
\newblock {\em Phys. Rev. E}, 85:047201, Apr 2012.

\bibitem{beenakker1997rmp}
C.~W.~J. Beenakker.
\newblock Random-matrix theory of quantum transport.
\newblock {\em Rev. Mod. Phys.}, 69:731--808, Jul 1997.

\bibitem{alhassid2000}
Y.~Alhassid.
\newblock The statistical theory of quantum dots.
\newblock {\em Rev. Mod. Phys.}, 72:895--968, Oct 2000.

\bibitem{bauer1990}
W.~Bauer and G.~F. Bertsch.
\newblock Decay of ordered and chaotic systems.
\newblock {\em Phys. Rev. Lett.}, 65:2213--2216, Oct 1990.
\newblock see also~\cite{legrand1991,bauer1991}.

\bibitem{richter2002}
Klaus Richter and Martin Sieber.
\newblock Semiclassical theory of chaotic quantum transport.
\newblock {\em Phys. Rev. Lett.}, 89:206801, Oct 2002.

\bibitem{waltner2008}
Daniel Waltner, Martha Guti\'errez, Arseni Goussev, and Klaus Richter.
\newblock Semiclassical mechanism for the quantum decay in open chaotic
  systems.
\newblock {\em Phys. Rev. Lett.}, 101:174101, Oct 2008.

\bibitem{raizen2011}
Mark Raizen and Daniel~A. Steck.
\newblock Cold atom experiments in quantum chaos.
\newblock {\em Scholarpedia}, 6(10):10468, 2011.

\bibitem{milner2001}
V.~Milner, J.~L. Hanssen, W.~C. Campbell, and M.~G. Raizen.
\newblock Optical billiards for atoms.
\newblock {\em Phys. Rev. Lett.}, 86(8):1514--1517, February 2001.

\bibitem{friedman2001}
Nir Friedman, Ariel Kaplan, Dina Carasso, and Nir Davidson.
\newblock Observation of chaotic and regular dynamics in atom-optics billiards.
\newblock {\em Phys. Rev. Lett.}, 86(8):1518--1521, February 2001.

\bibitem{kaplan2001}
Ariel Kaplan, Nir Friedman, Mikkel Andersen, and Nir Davidson.
\newblock Observation of islands of stability in soft wall atom-optics
  billiards.
\newblock {\em Phys. Rev. Lett.}, 87(27):274101, December 2001.

\bibitem{alt1996}
H.~Alt, H.-D. Gr{\"a}f, H.~L. Harney, R.~Hofferbert, H.~Rehfeld, A.~Richter,
  and P.~Schardt.
\newblock Decay of classical chaotic systems: The case of the {B}unimovich
  stadium.
\newblock {\em Phys. Rev. E}, 53(3):2217--2222, March 1996.

\bibitem{bloch2012nphysrev}
Immanuel Bloch, Jean Dalibard, and Sylvain Nascimb{\`e}ne.
\newblock Quantum simulations with ultracold quantum gases.
\newblock {\em Nat Phys}, 8(4):267--276, April 2012.

\bibitem{lewenstein2012book}
Maciej Lewenstein, Anna Sanpera, and Veronica Ahufinger.
\newblock {\em Ultracold Atoms in Optical Lattices: Simulating quantum
  many-body systems}.
\newblock Oxford University Press, Oxford, UK, May 2012.

\bibitem{politi2008}
Alberto Politi, Martin~J. Cryan, John~G. Rarity, Siyuan Yu, and Jeremy~L.
  {O'Brien}.
\newblock Silica-on-silicon waveguide quantum circuits.
\newblock {\em Science}, 320(5876):646--649, May 2008.

\bibitem{obrien2009}
Jeremy~L. {O'Brien}, Akira Furusawa, and Jelena Vuckovic.
\newblock Photonic quantum technologies.
\newblock {\em Nat. Photon.}, 3(12):687--695, December 2009.

\bibitem{haake}
Fritz Haake.
\newblock {\em Quantum Signatures of Chaos}.
\newblock Springer Series in Synergetics. Springer, Heidelberg, 2nd edition,
  2001.
\newblock chapters 4 and 5.

\bibitem{berry1977}
M.~V. Berry and M.~Tabor.
\newblock Level clustering in the regular spectrum.
\newblock {\em Proc. R. Soc. Lond. A}, 356(1686):375--394, September 1977.

\bibitem{bohigas1984}
O.~Bohigas, M.~J. Giannoni, and C.~Schmit.
\newblock Characterization of chaotic quantum spectra and universality of level
  fluctuation laws.
\newblock {\em Phys. Rev. Lett.}, 52(1):1--4, January 1984.

\bibitem{stockmann1990}
H.-J. St{\"o}ckmann and J.~Stein.
\newblock {‘‘Quantum’’} chaos in billiards studied by microwave
  absorption.
\newblock {\em Phys. Rev. Lett.}, 64(19):2215--2218, May 1990.

\bibitem{altland97}
Alexander Altland and Martin~R. Zirnbauer.
\newblock Nonstandard symmetry classes in mesoscopic normal-superconducting
  hybrid structures.
\newblock {\em Phys. Rev. B}, 55:1142--1161, Jan 1997.

\bibitem{montangero2009}
S.~Montangero, D.~Frustaglia, T.~Calarco, and R.~Fazio.
\newblock Quantum billiards in optical lattices.
\newblock {\em EPL (Europhysics Letters)}, 88(3):30006, 2009.

\bibitem{relano2002}
A.~Rela\~no, J.~M.~G. G\'omez, R.~A. Molina, J.~Retamosa, and E.~Faleiro.
\newblock Quantum chaos and $1/f$ noise.
\newblock {\em Phys. Rev. Lett.}, 89:244102, Nov 2002.

\bibitem{faleiro2004}
E.~Faleiro, J.~M.~G. G\'omez, R.~A. Molina, L.~Mu\~noz, A.~Rela\~no, and
  J.~Retamosa.
\newblock Theoretical derivation of $1/f$ noise in quantum chaos.
\newblock {\em Phys. Rev. Lett.}, 93:244101, Dec 2004.

\bibitem{faleiro2006}
E.~Faleiro, U.~Kuhl, R.~A. Molina, L.~Mu\~noz, A.~Rela\~no, and J.~Retamosa.
\newblock Power spectrum analysis of experimental sinai quantum billiards.
\newblock {\em Phys. Lett. A}, 358:251--255, Dec 2006.

\bibitem{dettmann2009}
Carl~P. Dettmann and Orestis Georgiou.
\newblock Survival probability for the stadium billiard.
\newblock {\em Physica D: Nonlinear Phenomena}, 238(23-24):2395--2403, 2009.

\bibitem{alt1995}
H.~Alt, H.~D. Gr\"af, H.~L. Harney, R.~Hofferbert, H.~Lengeler, A.~Richter,
  P.~Schardt, and H.~A. Weidenm\"uller.
\newblock {Gaussian Orthogonal Ensemble} statistics in a microwave stadium
  billiard with chaotic dynamics: {Porter-Thomas} distribution and algebraic
  decay of time correlations.
\newblock {\em Phys. Rev. Lett.}, 74:62--65, Jan 1995.

\bibitem{epaps}
The dynamical evolution from the initial state to the final lattice scar can be
  followed in the videos attached with this submission.

\bibitem{loeck2012}
Steffen L{\"o}ck, Arnd B{\"a}cker, and Roland Ketzmerick.
\newblock Coupling of bouncing-ball modes to the chaotic sea and their counting
  function.
\newblock {\em Phys. Rev. E}, 85:016210, Jan 2012.

\bibitem{inui1994}
M.~Inui, S.~A. Trugman, and Elihu Abrahams.
\newblock Unusual properties of midband states in systems with off-diagonal
  disorder.
\newblock {\em Phys. Rev. B}, 49:3190--3196, Feb 1994.

\bibitem{fleischer2003}
Jason~W. Fleischer, Mordechai Segev, Nikolaos~K. Efremidis, and Demetrios~N.
  Christodoulides.
\newblock Observation of two-dimensional discrete solitons in optically induced
  nonlinear photonic lattices.
\newblock {\em Nature}, 422(6928):147--150, March 2003.

\bibitem{krimer2011}
Dmitry~O. Krimer and Ramaz Khomeriki.
\newblock Realization of discrete quantum billiards in a two-dimensional
  optical lattice.
\newblock {\em Phys. Rev. A}, 84:041807, Oct 2011.

\bibitem{szameit2010jpb}
Alexander Szameit and Stefan Nolte.
\newblock Discrete optics in femtosecond-laser-written photonic structures.
\newblock {\em Journal of Physics B: Atomic, Molecular and Optical Physics},
  43(16):163001, 2010.

\bibitem{garanovich2012}
Ivan~L. Garanovich, Stefano Longhi, Andrey~A. Sukhorukov, and Yuri~S. Kivshar.
\newblock Light propagation and localization in modulated photonic lattices and
  waveguides.
\newblock {\em Physics Reports}, 518(1-2):1--79, 2012.
\newblock Light propagation and localization in modulated photonic lattices and
  waveguides.

\bibitem{corrielli2013}
Giacomo Corrielli, Andrea Crespi, Giuseppe {Della Valle}, Stefano Longhi, and
  Roberto Osellame.
\newblock Fractional {Bloch} oscillations in photonic lattices.
\newblock {\em Nature Communications}, 4:1555, 2013.

\bibitem{christo2003}
Demetrios~N. Christodoulides, Falk Lederer, and Yaron Silberberg.
\newblock Discretizing light behaviour in linear and nonlinear waveguide
  lattices.
\newblock {\em Nature}, 424:817--823, 2003.

\bibitem{ng2009}
G.~S. Ng, H.~Hennig, R.~Fleischmann, T.~Kottos, and T.~Geisel.
\newblock Avalanches of {Bose–Einstein} condensates in leaking optical
  lattices.
\newblock {\em New J. Phys.}, 11(7):073045, July 2009.

\bibitem{bakr2009}
Waseem~S. Bakr, Jonathon~I. Gillen, Amy Peng, Simon Fölling, and Markus
  Greiner.
\newblock A quantum gas microscope for detecting single atoms in a
  hubbard-regime optical lattice.
\newblock {\em Nature}, 462(7269):74--77, November 2009.

\bibitem{wurtz2009}
Peter W\"urtz, Tim Langen, Tatjana Gericke, Andreas Koglbauer, and Herwig Ott.
\newblock Experimental demonstration of single-site addressability in a
  two-dimensional optical lattice.
\newblock {\em Phys. Rev. Lett.}, 103(8):080404, August 2009.

\bibitem{sherson2010}
Jacob~F. Sherson, Christof Weitenberg, Manuel Endres, Marc Cheneau, Immanuel
  Bloch, and Stefan Kuhr.
\newblock Single-atom-resolved fluorescence imaging of an atomic mott
  insulator.
\newblock {\em Nature}, 467(7311):68--72, September 2010.

\bibitem{morozov2006}
S.~V. Morozov, K.~S. Novoselov, M.~I. Katsnelson, F.~Schedin, L.~A.
  Ponomarenko, D.~Jiang, and A.~K. Geim.
\newblock Strong suppression of weak localization in graphene.
\newblock {\em Phys. Rev. Lett.}, 97:016801, Jul 2006.

\bibitem{dassarma2011rmp}
S.~Das~Sarma, Shaffique Adam, E.~H. Hwang, and Enrico Rossi.
\newblock Electronic transport in two-dimensional graphene.
\newblock {\em Rev. Mod. Phys.}, 83:407--470, May 2011.

\bibitem{houck2012}
D.~L. Underwood, W.~E. Shanks, Jens Koch, and A.~A. Houck.
\newblock Low-disorder microwave cavity lattices for quantum simulation with
  photons.
\newblock {\em Phys. Rev. A}, 86:023837, Aug 2012.

\bibitem{Peropadre2013}
Borja Peropadre, David Zueco, Friedrich Wulschner, Frank Deppe, Achim Marx,
  Rudolf Gross, and Juan~Jos\'e Garc{\'\i}a-Ripoll.
\newblock Tunable coupling engineering between superconducting resonators: From
  sidebands to effective gauge fields.
\newblock {\em Phys. Rev. B}, 87:134504, Apr 2013.

\bibitem{plenio2008}
M.~B. Plenio and S.~F. Huelga.
\newblock Dephasing-assisted transport: quantum networks and biomolecules.
\newblock {\em New J. Phys.}, 10(11):113019, November 2008.

\bibitem{seeman2005}
Nadrian~C Seeman and Philip~S Lukeman.
\newblock Nucleic acid nanostructures: bottom-up control of geometry on the
  nanoscale.
\newblock {\em Reports on Progress in Physics}, 68(1):237, 2005.

\bibitem{mello2004}
Pier~A. Mello and Narendra Kumar.
\newblock {\em Quantum Transport in Mesoscopic Systems. Complexity and
  Statistical Fluctuations}, volume~4 of {\em Mesoscopic Physics and
  Nanotechnology}.
\newblock Oxford University Press, Oxford, 2004.

\bibitem{datta2005}
Supriyo Datta.
\newblock {\em Quantum Transport: Atom to Transistor}.
\newblock Cambridge University Press, Cambridge, UK, 2005.

\bibitem{perets2008}
Hagai~B. Perets, Yoav Lahini, Francesca Pozzi, Marc Sorel, Roberto Morandotti,
  and Yaron Silberberg.
\newblock Realization of quantum walks with negligible decoherence in waveguide
  lattices.
\newblock {\em Phys. Rev. Lett.}, 100(17):170506, May 2008.

\bibitem{karski2009}
Michal Karski, Leonid F\"orster, Jai-Min Choi, Andreas Steffen, Wolfgang Alt,
  Dieter Meschede, and Artur Widera.
\newblock Quantum walk in position space with single optically trapped atoms.
\newblock {\em Science}, 325(5937):174--177, July 2009.
\newblock {PMID:} 19589996.

\bibitem{broome2013}
Matthew~A. Broome, Alessandro Fedrizzi, Saleh Rahimi-Keshari, Justin Dove,
  Scott Aaronson, Timothy~C. Ralph, and Andrew~G. White.
\newblock Photonic boson sampling in a tunable circuit.
\newblock {\em Science}, 339(6121):794--798, February 2013.
\newblock {PMID:} 23258411.

\bibitem{spring2013}
Justin~B. Spring, Benjamin~J. Metcalf, Peter~C. Humphreys, W.~Steven
  Kolthammer, Xian-Min Jin, Marco Barbieri, Animesh Datta, Nicholas
  Thomas-Peter, Nathan~K. Langford, Dmytro Kundys, James~C. Gates, Brian~J.
  Smith, Peter G.~R. Smith, and Ian~A. Walmsley.
\newblock Boson sampling on a photonic chip.
\newblock {\em Science}, 339(6121):798--801, February 2013.
\newblock {PMID:} 23258407.

\bibitem{tillmann2013}
Max Tillmann, Borivoje Dakić, René Heilmann, Stefan Nolte, Alexander Szameit,
  and Philip Walther.
\newblock Experimental boson sampling.
\newblock {\em Nat Photon}, 7(7):540--544, July 2013.

\bibitem{crespi2013}
A.~Crespi, R.~Osellame, R.~Ramponi, D.~J. Brod, E.~F. Galvao, N.~Spagnolo,
  C.~Vitelli, E.~Maiorino, P.~Mataloni, and F.~Sciarrino.
\newblock Integrated multimode interferometers with arbitrary designs for
  photonic boson sampling.
\newblock {\em Nature Photonics}, 7(7):545--549, May 2013.

\bibitem{Porter56}
C.~E. Porter and R.~G. Thomas.
\newblock Fluctuations of nuclear reaction widths.
\newblock {\em Phys. Rev.}, 104:483--491, Oct 1956.

\bibitem{Brody81}
T.~A. Brody, J.~Flores, J.~B. French, P.~A. Mello, A.~Pandey, and S.~S.~M.
  Wong.
\newblock Random-matrix physics: spectrum and strength fluctuations.
\newblock {\em Rev. Mod. Phys.}, 53:385--479, Jul 1981.

\bibitem{Berry77}
M.V. Berry.
\newblock Regular and irregular semiclassical wave functions.
\newblock {\em J. Phys. A}, 10:2083--2091, 1977.

\bibitem{Meredith93}
D.~C. Meredith.
\newblock Statistics and scarring of eigenvectors of a shell model.
\newblock {\em Phys. Rev. E}, 47:2405--2414, Apr 1993.

\bibitem{legrand1991}
Olivier Legrand and Didier Sornette.
\newblock First return, transient chaos, and decay in chaotic systems.
\newblock {\em Phys. Rev. Lett.}, 66:2172--2172, Apr 1991.

\bibitem{bauer1991}
Wolfgang Bauer and George~F. Bertsch.
\newblock Bauer and {B}ertsch reply.
\newblock {\em Phys. Rev. Lett.}, 66:2173--2173, Apr 1991.

\end{thebibliography}

\end{document}